%% file: main.tex
\documentclass{article}
\usepackage{waspaa21,amsmath,graphicx,url,times}
\usepackage{amssymb, multirow, booktabs}
\usepackage[ruled,vlined]{algorithm2e}
\usepackage{color}

\SetKwInput{KwInput}{Input}  
\SetKwInput{KwOutput}{Output}

\title{Sound Event Detection with Adaptive Frequency Selection}

\name{Zhepei Wang,$^\sharp$ \thanks{\texttt{https://github.com/zhepeiw/adaptive\_freq\_select}}
      Jonah Casebeer,$^\sharp$
      Adam Clemmitt,$^\sharp$
      Efthymios Tzinis,$^\sharp$
      Paris Smaragdis,$^\sharp$$^\flat$}
\address{$\sharp$ University of Illinois at Urbana-Champaign, Computer Science Dept., \\              
         $\flat$ Adobe Research \\
}

\begin{document}
\input{commands}

\ninept
\maketitle

\begin{sloppy}

\begin{abstract}
  In this work, we present HIDACT, a novel network architecture for adaptive computation for efficiently recognizing acoustic events. We evaluate the model on a sound event detection task where we train it to adaptively process frequency bands. The model learns to adapt to the input without requesting all frequency sub-bands provided. It can make confident predictions within fewer processing steps, hence reducing the amount of computation. Experimental results show that HIDACT has comparable performance to baseline models with more parameters and higher computational complexity. Furthermore, the model can adjust the amount of computation based on the data and computational budget.
  
  %Deep neural networks have achieved significant progress for sound event detection tasks. However, these models often contain millions of parameters and have high computational complexity. With sub-band processing and adaptive frequency selection, we propose an efficient sound event detection system with significantly smaller model size as well as computational complexity. Our model learns to adapt to the input and intelligently select the number of frequency sub-bands, making a confident prediction within fewer processing steps. Furthermore, our model can adjust the amount of computation based on the data and computational budget. Results show that the proposed method has comparable performance to baseline models with higher complexity on monophonic event detection tasks.
% Thymios: I think this needs rephrasing, you do not convey the full potential of HIDACT.  
\end{abstract}


\input{intro}

\input{method}

\input{expr}

\begin{figure*}[!htb]
  \centering
  \includegraphics[width=0.92\linewidth, height=0.18\textheight]{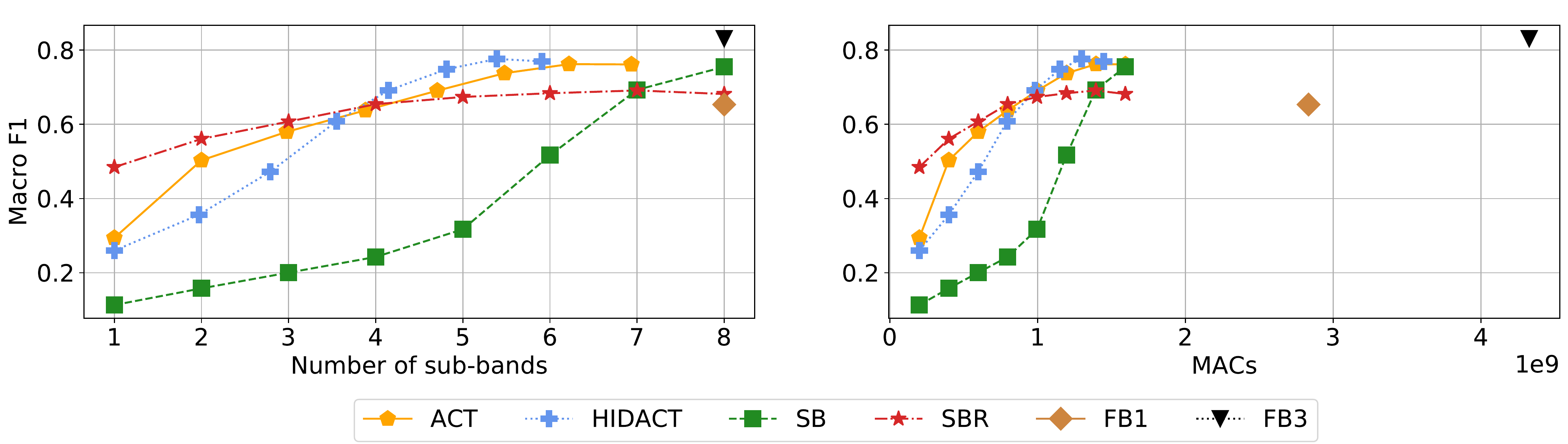}
\caption{Test macro f1 scores for the proposed adaptive computation and baseline systems.  The proposed adaptive methods achieve comparable f1 scores within significantly fewer steps and MACs than the baseline models.}
\label{fig:f1_multiclass}
\end{figure*}

% removed caption: The left figure shows the f1 against the maximum amount of input sub-bands the model observes; the right figure shows the f1 versus the number of MACs.

\begin{figure}[!htb]
  \centering
  \includegraphics[width=0.85\linewidth]{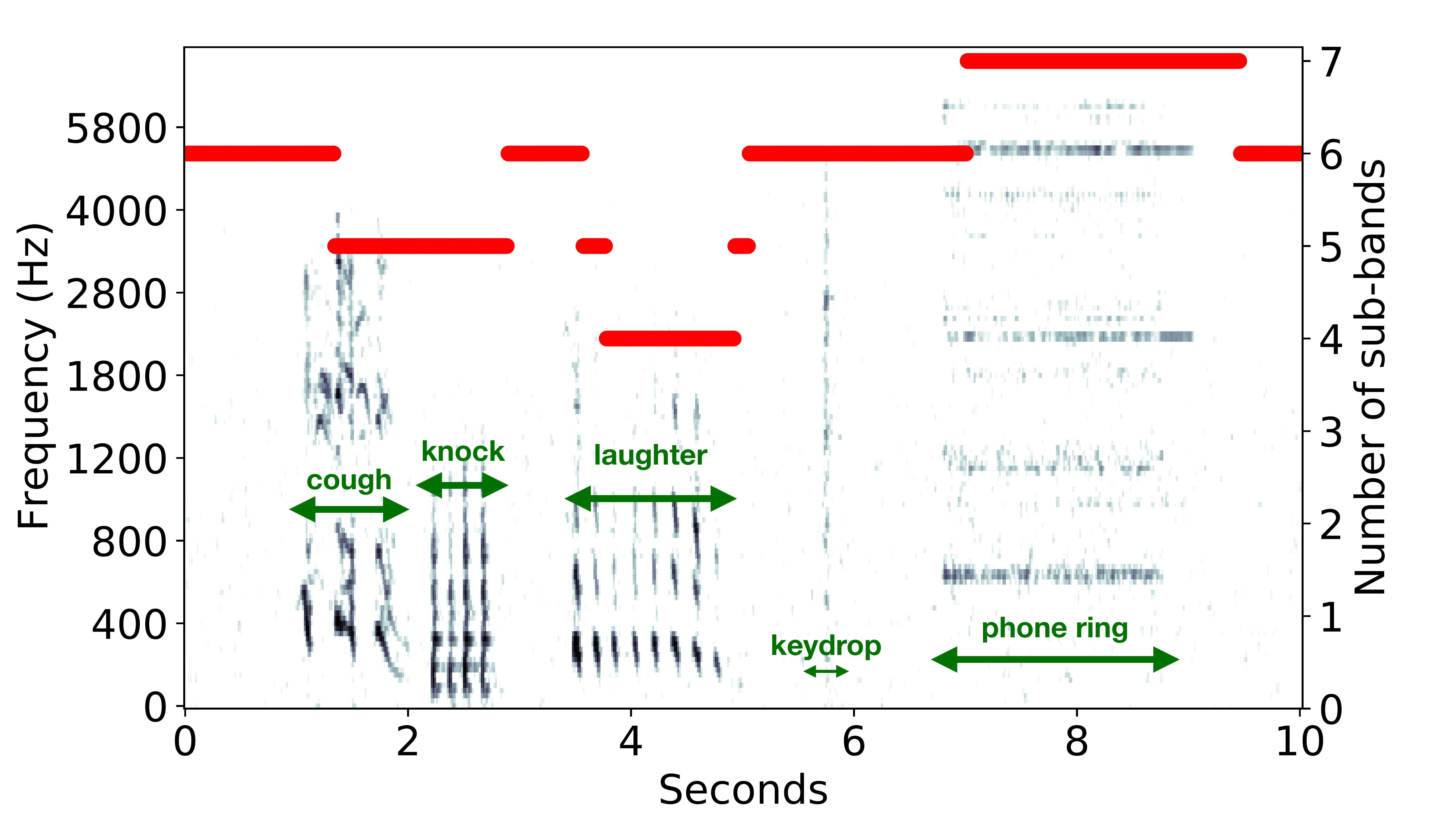}
\caption{The number of sub-bands processed by the HIDACT model for a test recording overlayed on the log mel-spectrogram. We include arrows to indicate the ground-truth event labels. HIDACT can adjust the number of sub-bands it requests and processes based on the spectral properties of the input signal.}
\label{fig:hidact_step}
\end{figure}
\input{results}

\input{conclusion}

% -------------------------------------------------------------------------
% Either list references using the bibliography style file IEEEtran.bst
\bibliographystyle{IEEEtran}
\bibliography{refs21}
%
% or list them by yourself
% \begin{thebibliography}{9}
% 
% \bibitem{waspaa21web}
%   \url{http://www.waspaa.com}.
%
% \bibitem{IEEEPDFSpec}
%   {PDF} specification for {IEEE} {X}plore$^{\textregistered}$,
%   \url{http://www.ieee.org/portal/cms_docs/pubs/confstandards/pdfs/IEEE-PDF-SpecV401.pdf}.
%
% \bibitem{PDFOpenSourceTools}
%   Creating high resolution {PDF} files for book production with 
%   open source tools, 
%   \url{http://www.grassbook.org/neteler/highres_pdf.html}.
%
% \bibitem{eWilliams1999}
% E. Williams, \emph{Fourier Acoustics: Sound Radiation and Nearfield Acoustic
%   Holography}. London, UK: Academic Press, 1999.
% 
% \bibitem{ieeecopyright}
%   \url{http://www.ieee.org/web/publications/rights/copyrightmain.html}.
%
% \bibitem{cJones2003}
% C. Jones, A. Smith, and E. Roberts, ``A sample paper in conference
%   proceedings,'' in \emph{Proc. IEEE ICASSP}, vol. II, 2003, pp. 803--806.
% 
% \bibitem{aSmith2000}
% A. Smith, C. Jones, and E. Roberts, ``A sample paper in journals,'' 
%   \emph{IEEE Trans. Signal Process.}, vol. 62, pp. 291--294, Jan. 2000.
% 
% \end{thebibliography}

\end{sloppy}
\end{document}

%% file: commands.tex
\newcommand{\R}{\mathbb{R}}
\newcommand{\ba}{\mathbf{a}}
\newcommand{\bh}{\mathbf{h}}
\newcommand{\bq}{\mathbf{q}}
\newcommand{\bs}{\mathbf{s}}
\newcommand{\bx}{\mathbf{x}}
\newcommand{\by}{\mathbf{y}}

%% file: intro.tex
\section{Introduction}
\label{sec:intro}

% Recent improvement in learning-based models for sound event detection is tightly associated with larger and deeper neural network architectures. Convolutional recurrent network (CRNN) \cite{crnn_2017} has obtained great success in audio classification tasks, where a stack of convolutional layers are first applied to learn local representation of the input features followed by one or more recurrent layers to model temporal information. Most leading systems of the sound event detection tasks for the Detection and Classification of Acoustic Scenes and Events (DCASE) in recent years follow the design paradigm, with variants of convolutional blocks \cite{CRNN_DCASE_2018, CRNN_DCASE_2019} or using Transformers in place of recurrent neural networks (RNN)s \cite{C_Transformer_DCASE_2020}.

Deep neural networks have achieved significant progress for a variety of audio processing tasks. Large models with millions of parameters have delivered new state of the art results in sound event detection \cite{crnn_taslp}, scene classification \cite{ASC_subband_CNN_DCASE2020}, audio captioning \cite{DCASE_2020_caption}, and more \cite{pann}. The convolutional recurrent neural network~(CRNN) \cite{crnn_taslp}, composed of a series of convolutional and recurrent layers, has become a dominant architecture in event detection competitions \cite{DCASE_2020_Task4}. The success of the CRNN architecture can be attributed to its ability to combine local information with convolutional layers and temporal information with recurrent layers. Several promising lines of work are based on improving these structures \cite{C_Transformer_DCASE_2020, CRNN_DCASE_2018, CRNN_DCASE_2019}. However, the deployment of these deep neural networks is complex as different devices have different computational, storage, or bandwidth limits.

 To address this issue, multiple concurrent lines of work have proposed techniques to reduce the complexity of the model for audio processing tasks. These techniques include sparsity-based approaches \cite{han2015deep, han2015learning, liu2015sparse, casebeer2020efficient}, and quantization based approaches \cite{bnn_sed, bnn_ssp, brnn_ssp}. In addition, a variety of weight-sharing approaches have demonstrated strong results. Sub-band processing, an approach that shares parameters across bands of an input, has successfully decreased computational costs while maintaining strong performance in sound event detection \cite{subband_dcase_2018}, acoustic scene classification \cite{ASC_subband_CNN_DCASE2020}, speech recognition \cite{FLSTM, subband_spoken_term}, speech enhancement \cite{subband_se} and source separation tasks \cite{groupcomm, gc3}. However, these methods typically assume that models have access to a fixed number of bands or inputs.

In scenarios where the number of input channels or bands is time-varying, systems must scale to large or small numbers of inputs. The multi-view network~(MVN) provides a way to train a model on a fixed number of input channels and test on a varying number of inputs \cite{mvn_se, mvn_clf}. Often, the number of inputs is determined by the computational budget. Therefore, adaptive computation models have been proposed to minimize computation in a time-varying fashion. These models also incorporate computational expense into their training objectives. Adaptive computation time~(ACT), originally proposed by \cite{ACT}, achieves this by constructing weighted sums of hidden states and penalizing excessive hidden state usage. The ACT framework has been successfully applied to a variety of other tasks including early exiting of convolutional neural network~(CNN) layers \cite{branchynet, sact}, adaptive depth Transformers \cite{adaptive_trf}, microphone selection \cite{casebeer2020communication}, and in visual question answering through the differentiable adaptive computation time~(DACT) mechanism \cite{DACT}.

% With sub-band processing and adaptive computation, we propose a system with low computational complexity that performs sound event detection without overlapping events. We experimentally show that the proposed method has a comparable performance to other baseline models that process all frequency sub-bands with higher complexity or larger model size. We also present visually that the proposed model adapts well to the input and makes varying number of processing steps depending on the content of the signal.

In this work, we propose a technique to train models with a fixed computational budget and deploy them with an intelligently adaptive budget. We demonstrate this technique in a monophonic sound event detection task where the proposed models vary the number of frequency-bins requested across time. We compare these models to a suite of baselines, including full-band models and sub-band models observing all frequency sub-bands. To verify that the proposed models are compatible with the aforementioned advanced neural network architectures for event detection, we integrate the network design and learning objectives of the adaptive computation with the CRNN architecture in \cite{crnn_taslp, CRNN_DCASE_2018, CRNN_DCASE_2019}, as it is the most popular architecture in recent DCASE challenges \cite{DCASE_2020_Task4}. We evaluate the classification performance with frame-wise macro F1. We also evaluate the computational cost of all techniques with the multiply-accumulate (MAC) operations \cite{macs}. Empirical results show that these adaptive models can gracefully scale their computational budgets effectively, allowing a single model to be trained and then deployed on devices with a varying computational budget.

% --- backup text
%These rapid gains in performance have been accompanied by a rapid increase in computational and storage complexity. This complexity has hindered the deployment of deep learning models to edge, mobile, or other lightweight devices.

%% file: method.tex
\section{System Descriptions}
\label{sec:method}

\subsection{Dynamic Processing with Multi-View Networks}
\label{ssec:mvn}

MVNs are initially proposed for multi-channel speech enhancement \cite{mvn_se} and voice activity detection tasks \cite{mvn_clf} with a potentially variable number of channels across time. In this work, we focus on single-channel audio input where each feature corresponds to a frequency sub-band. Unrolling across both feature and temporal dimensions, MVNs can process a time-varying number of steps.

For each frame, the network observes all information before making a prediction. Then, the hidden state of the last step of the current frame is passed to the processing of the first feature of the next frame, allowing the propagation of temporal information. The number of features at one frame is potentially different from another, and the network is capable of dynamically changing the number of processing steps over time. For a given time step $t$ and feature unrolling step $k$, the recurrence relation of the MVN can be written as 
\begin{align}
    \bs_t^k = \begin{cases}
    f_s(\bs_{t}^{k-1}, \bx_{t}^k), \quad & \text{if } k > 1, \\
    f_s(\bs_{t-1}, \bx_{t}^1), \quad & \text{if } k = 1, \\
    \end{cases}
    \label{eq:mvn_recurrence}
\end{align}
where $f_s$ is the recurrent network and $\bx_{t}^k \in \R^{d}$ is the $k$-th input feature at time $t$. $\bs_{t}^k$ refers to the hidden representation at frame $t$ and step $k$, while $\bs_t$ is the aggregate hidden state of frame $t$ and is chosen to be the output of the last feature $\bs_t = \bs_t^K$ in \cite{mvn_se, mvn_clf}, where $K$ is the total number of features at this frame. The output at this frame is $\by_t = f_y(\bs_t)$ 
% \begin{align}
%     \by_t = f_y(\bs_t),
%     \label{eq:mvn_out}
% \end{align}
where $f_y$ is usually a linear transformation.

\subsection{Adaptive Computation}
\label{ssec:adaptive}

MVNs are \textit{dynamic} but not \textit{adaptive}: the number of processing steps changes over time, but MVNs always make use of all available features (sub-bands) at each frame regardless of the input. To further reduce the amount of computation, it is desirable to have an \textit{adaptive} system which intelligently selects the number of frequency sub-bands for each frame instead of processing all sub-bands. 
%As a simple thought experiment, suppose there is an audio recording of speech followed by music. Intuitively, the low frequency bins of the speech frames have richer content, whereas the high frequency bins of the music frames contain more information. For each frame, the model unrolls from low frequency subbands to high frequency subbands. If the model is trained in a way that it is confident about the prediction based on the information it has seen, it can intelligently stop processing the current frame and continue to the next frame.
The model starts from the lowest frequency sub-bands and absorbs one sub-band at a time. It intelligently stops and begins the processing of the next frame when the implicit confidence model predicts that there is no significant gain for including more sub-bands.

\subsubsection{Adaptive Computation Time}
\label{sssec:act}
ACT \cite{ACT} is first proposed to train a recurrent network that can adjust the amount of computation for each step. With the hidden state $\bs_t^k$, it learns a computational cost $h_t^k = \sigma(f_h(\bs_t^k)) \in [0, 1]$, where $\sigma$ denotes the element-wise sigmoid function and $f_h$ is a learned linear transformation. ACT halts the computation at time $t$ for $N(t)$ steps once it exceeds the computational budget, as shown in next: 
\begin{align}
    N(t) = \min \{ N: \sum_{k=1}^N h_t^k \geq 1 - \epsilon \},
    \label{eq:act_n}
\end{align}
where $\epsilon$ is a small constant (i.e., 0.01), and the remainder is defined as \begin{align} 
    R(t) = 1 - \sum_{k=1}^{N(t)-1} h_t^n.
    \label{eq:act_r}
\end{align}
In addition to the task loss, the network is optimized with the ponder cost
\begin{align}
    \mathcal{L}_{act} = \sum_t N(t) + R(t),
    \label{eq:act_loss}
\end{align}
where minimizing the term $R(t)$ encourages the network to produce larger values of computational cost $h_t^k$ and hence to reduce the number of computational steps.

The aggregate hidden state $\bs_t$ and output $\by_t$ at frame $t$ are given by the weighted sum of the output of individual steps as follows:
\begin{align}
    \bs_t  = \sum_{k=1}^{N(t)} w_t^k \bs_t^k, \quad \by_t  = \sum_{k=1}^{N(t)} w_t^k \by_t^k,
    \label{eq:act_aggregation}
\end{align}
where the weights $w_t^k$ follow \begin{align}
    w_t^k = \begin{cases}
    h_t^k, \quad & \text{if } k < N(t), \\
    R(t), \quad & \text{if } k = N(t). \\
    \end{cases}
    \label{eq:act_w}
\end{align}
$\bs_t$ is passed to the processing of frame $t+1$. %Notice that $w_t^k \in [0, 1]$ and $\sum_{k=1}^{N(t)} w_t^k = 1$ is a valid probability distribution. 

\subsubsection{Differentiable Adaptive Computation Time}
\label{sssec:dact}
ACT suffers from instability in the number of processing steps and requires extensive hyperparameter tuning for good performance \cite{ casebeer2020communication, DACT}. This issue is attributed to the fact that the ACT pipeline is not fully differentiable. As described in (\ref{eq:act_r}) and (\ref{eq:act_w}), ACT imposes the constraint that the intermediate weights $w_t^k$ should sum to one. If the sum is greater than one, then the weight of the final step is clipped, making the final output non-differentiable.

To overcome this issue, Differentiable Adaptive Computation Time (DACT) \cite{DACT} is proposed as an end-to-end differentiable pipeline based upon ACT. It applies to halt only at test time while always processing all information during training. This is achieved by constructing the uncertainty $g_t^k = f_g(\bs_t^k) \in [0,1]$ at each step $k$, where $f_g$ is the network. The probability that a subsequent processing step $k' > k$ may change the model's prediction is defined to be the product $p_t^k = \prod_{i=1}^k g_t^k\in [0,1]$. The ponder cost is defined as \begin{align}
    \mathcal{L}_{dact} = \sum_t \sum_k p_t^K,
    \label{eq:dact_loss}
\end{align}
where $K$ is the maximum number of processing steps. It encourages the model to be confident about its prediction at earlier steps.

The accumulated output $\ba_t^k$ is a linear combination of the step-wise output $\by_t^i$ from all previous processing steps $i\leq k$: \begin{align}
    \ba_t^k = \begin{cases}
    \mathbf{0}, \quad \text{if } k = 0, \\
    \by_t^k p_t^{k-1} + \ba_t^{k-1} (1 - p_t^{k-1}), \quad \text{if } k > 0. \\
    \end{cases}
    \label{eq:dact_a}
\end{align}

 Notice that if $g_t^k \rightarrow 0$ at step $k$ then $p_t^{k'} \rightarrow 0$ for all subsequent steps $k' > k$; the model's prediction $\ba_t^k$ is hence unlikely to change. This property allows us to define an early stopping condition with a lower bound of $l^k[c^*]$ for the class $c^*$ with the highest probability and an upper bound $u^k[c']$ for the runner-up class $c'$ (details in Algorithm~\ref{alg:hidact}). During inference, we reach the halting condition once $l^k[c^*] > u^k[c']$. A theoretical proof of the bounds and the fact that the class label is guaranteed not to change by subsequent processing steps is presented in \cite{DACT}.
 
%  When the output $\ba^k$ follows a softmax distribution of $C$ classes, we can derive a lower bound $l^k[c^*]$ for the class $c^*$ with the highest probability and an upper bound $u^k[c']$ for the runner-up class $c'$:
%  \begin{align}
%      l^k [c^*] & = \ba^k [c^*] (1 - p^k)^{K-k}, \\
%      u^k [c'] &= \ba^k [c'] + p^k(K-k).
%      \label{eq:dact_bound}
%  \end{align}
% During inference, we reach the halting condition once $l^k[c^*] > u^k[c']$. A theoretical proof of the bounds and the fact that the class label is guaranteed not to change by subsequent processing steps is presented in \cite{DACT}.

% % TODO: HIDACT

% \subsubsection{Hidden State-Interpolated Differentiable Adaptive Computation Time}
% \label{sssec:hidact}

DACT is initially proposed for processing non-temporal information and is not concerned with passing hidden states between temporal steps. Hence, it does not perform interpolation in the hidden space. When unrolling across both spectral and temporal dimensions, the hidden state at the last step of each frame feeds into the processing of the next frame. Preliminary experiments show that this formulation leads to significant degradation in test performance. One possible explanation is that the model always processes the maximum number of steps $K$ during training, but the termination condition might be reached earlier during inference with a hidden state distinct from processing all $K$ steps.

We overcome this problem with the proposed Hidden-state Interpolated DACT (HIDACT) by interpolating between the hidden states similar to the interpolation of the aggregate output in (\ref{eq:dact_a}). We describe the process of constructing the aggregate hidden state $\bq_t$ and output $\ba_t$ at each step in Algorithm~\ref{alg:hidact}. With this interpolation, we impose a constraint on the accumulated hidden state such that it cannot change significantly once the uncertainty $p_t^k$ becomes small and the halt criterion is met, significantly improving the performance and stability during test time.

% Similar to (\ref{eq:dact_a}) we define the accumulated hidden state $\bq_t^k$ as the weighted sum of the step-wise hidden state $\bs_t^k$: \begin{align}
%     \bq_t^k = \begin{cases}
%     \mathbf{0}, \quad \text{if } k = 0, \\
%     \bs_t^k p_t^{k-1} + \bq_t^{k-1} (1 - p_t^{k-1}), \quad \text{if } k > 0. \\
%     \end{cases}
%     \label{eq:dact_q}
% \end{align}

% During training, the model observes all $K$ steps for each frame. The accumulated hidden state of the last step $\bq_t^K$ at frame $t$ feeds into the processing of frame $t+1$. During inference, if the halting condition is met at step $i < K$, then the accumulated hidden state $\bq_t^i$ is passed to the next frame.

% TODO: figure of HIDACT

\begin{algorithm}[htb!]
\SetAlgoLined
\KwInput{$f_s$, $f_g$, $f_y$, $\bx \in \R^{T\times K}$}
\KwOutput{$\ba \in \R^{T\times C}$, $\mathcal{L} $}
$\mathcal{L} \gets 0$ \\
\For{$t \gets 1$ to $T$} {
    \tcp{Aggregate quantities}
    $\bq_t \gets 0 $; $\ba_t \gets 0 $; $p  \gets 1 $ \\
    \For{$k \gets 1$ to $K$}{
        \uIf{$k == 1$} {
            $\bs_t^k  \gets  f_s(\bq_{t-1}, \bx_t^k)$
        }
        \Else {
        $\bs_t^k  \gets  f_s(\bs_t^{k-1}, \bx_t^k)$
        }
        $\by_t^k \gets f_y(\bs_t^k)$ \\
        $g_t^k \gets f_g(\bs_t^k)$ \\
        \tcp{Update aggregate quantities}
        $\bq_t \gets \bs_t^k \cdot p + \bq_t \cdot (1 - p)$ \\
        $\ba_t \gets \by_t^k \cdot p + \ba_t \cdot (1 - p)$ \\
        $p \gets g_t^k \cdot p$ \\
        \uIf{training}{
            \tcp{Update loss}
            $\mathcal{L} \gets \mathcal{L} + p$
        }
        \Else{
            \tcp{Check halting condition}
            $c_{\text{sort}} \gets \text{argsort}_c \ba_t[c] $ \tcp{Descending}
            $c^* \gets c[1]$; $c' \gets c[2]$ \\
            $l[c^*] \gets \ba_t[c^*](1-p)^{K-k}$ \\
            $u[c'] \gets \ba_t[c'] + p\cdot (K-k)$ \\
            \uIf{$l[c^*] > u[c']$}{
                break \\
            }
        }

    }
}

\caption{Forward pass of HIDACT.}
 \label{alg:hidact}
\end{algorithm}

%% file: expr.tex
\section{Experimental Setup}
\label{sec:expr}

% We apply the proposed adaptive frequency selection method on sound event detection tasks where the audio input does not contain overlapping events. This is in essence a multi-class classification problem. The following sections explain the dataset, models and training configurations for our experiments.

\subsection{Dataset}
\label{ssec:expr_data}

We synthesize a dataset for monophonic sound event detection due to the limited availability of annotated data without overlapping events. Our data is synthesized from the dataset of DCASE 2016 Task 2 \cite{dcase_2016}. The training split of the DCASE dataset consists of 20 isolated segments for each one of the 11 event classes. We randomly select 16 segments from each class to synthesize our training set and 4 segments for the test set. These segments are downsampled to 16k Hz and mixed with Gaussian noise
%noise segments selected from 13 different background noise recordings such as \textit{air-port}, \textit{babble} and \textit{restaurant} noises \cite{noise-16k}
with the signal-to-noise ratio (SNR) at 0 dB. Each one of the synthesized recordings is 10 seconds long, and the training set contains 1000 recordings while the test set contains 200 recordings. We compute the log mel-spectrogram with 128 features with a window size of 1024 and a hop size of 256. 

\subsection{Model Configuration}
\label{ssec:expr_model}

 We partition the spectral frames into 8 non-overlapping sub-bands so that each sub-band contains 16 contiguous frequency bins. Similar to the architecture in \cite{CRNN_DCASE_2018, CRNN_DCASE_2019}, our sub-band network contains four convolutional blocks with 16, 32, 64, and 128 channels, respectively, followed by a 128-dimensional unidirectional GRU with softmax output heads. We fix the kernel size to be $1\times 1$ to reduce the computation cost, and we perform convolution within each frequency sub-band.

% \subsubsection{HIDACT and ACT}
% \label{sssec:expr_proposed}

% The proposed system consists of sub-band processing and adaptive computation to intelligently select the number of frequency sub-bands required during inference.
%We also experiment with seven-block CNNs with three additional blocks with 128 channels appended to the four-block CNN.
% To reduce repeated processing of neighboring frames, we fix the kernel sizes to be $1\times 1$ for all convolutional layers so that when processing the sub-band at the current frame the model does not leverage additional information from other frames, and the temporal dependency is processed by the GRU. The system is trained with the objective function \begin{align}
%     \mathcal{L} = \mathcal{L}_{ce} + \lambda \mathcal{L}_p,
%     \label{eq:system_loss}
% \end{align}
% where $\mathcal{L}_{ce}$ is the cross-entropy loss and $\mathcal{L}_p$ defined in (\ref{eq:act_loss}) or (\ref{eq:dact_loss}) is the ponder loss. $\lambda$ is a hyperparameter controlling the weights of the ponder term, and is set to 1e-3 in our experiments.

We compare the proposed method with several sub-band and full-band baseline models. The sub-band baseline model (SB) has the same model configuration as the proposed system with the exception that the baseline is trained with only the cross-entropy loss $\mathcal{L}_{ce}$. We also propose a variant of this baseline by masking out a uniformly random number (between 0 and 7) of contiguous frequency sub-bands during training (SBR).

The full-band model (FB1) has the same configuration as the sub-band model except that the GRU contains 2 layers, each with 512 hidden units, to handle the larger input dimension. We also experiment with a version of the full-band model using $3\times 3$ convolutional kernels (FB3). The full-band models are also trained with the cross-entropy loss only.

\begin{table}[!htb]
\centering
\begin{tabular}{c|c|c|c|c}
\hline
Name & \# Params & \# Training & \# Test & Loss \\
 & (1e6) &  bands &  bands & type \\
 \hline
 FB1 & 2,633  & \multirow{2}{*}{8} & \multirow{2}{*}{8} & \multirow{2}{*}{$\mathcal{L}_{ce}$}\\
 FB3 & 2,893 &  & \\
 \hline
 SB & \multirow{4}{*}{133} & 8 & \multirow{2}{*}{8} & \multirow{2}{*}{$\mathcal{L}_{ce}$} \\
 SBR &  & random &  & \\
 \cline{1-1} \cline{3-5}
 HIDACT &  & 8 & \multirow{2}{*}{adaptive} & $\mathcal{L}_{ce} + \lambda\mathcal{L}_{dact}$\\
 ACT &  & adaptive &  & $\mathcal{L}_{ce} + \lambda\mathcal{L}_{act}$ \\
\hline

\end{tabular}
\caption{Overview of the proposed and baseline systems.}
\label{tab:sys}
\end{table}

Table~\ref{tab:sys} shows the comparison between the proposed and each of the baseline models, including the model size, the number of sub-bands processed during training and inference, and the loss type. All sub-band systems have roughly the same number of parameters, and they only differ in the processing of intermediate hidden state and output. The sub-band models have significantly fewer parameters compared to the full-band models. All the baseline models except SBR process a fixed number of frequency sub-bands during both training and test time, while SBR is designed to process a random number of sub-bands during training but also makes a fixed amount of computation at test time. ACT is adaptive to the input during both training and test, while HIDACT always processes all sub-bands during training and adaptively selects the number of sub-bands during inference. All baseline models are trained only with the cross-entropy loss $\mathcal{L}_{ce}$, while the adaptive systems have an additional pondering cost. $\lambda$ is a hyperparameter controlling the weights of the ponder term and is chosen as 0.001 for ACT and 0.01 for HIDACT in our experiments.

\subsection{Training and Evaluation Details}
\label{ssec:expr_train}

During training, we randomly select a 2-second window (128 frames) from the entire input to speed up processing. All models are trained with a batch size of 16 and optimized with Adam \cite{adam} using an initial learning rate of 1e-4. We train the networks for a maximum of 100,000 iterations (roughly 1500 epochs).%, and we apply early stopping if the validation macro-f1 has not improved for 500 epochs.

We evaluate the model's performance on the 200 test recordings using the frame-wise macro F1 score of the 12-way classification, including the background class. The macro F1 score is defined as the mean of the class-wise F1 score. We use the amount of MAC operations \cite{macs} as a metric for computational complexity.

%% file: results.tex
\section{Results and Discussions}
\label{sec:res}

% \subsection{Architecture Search}
% \label{ssec:res_arch}
% We first perform a study on the performance of different choices of architecture for the adaptive computation methods ACT and HIDACT. We experiment with three different architectures: 1) a network with only a GRU layer, 2) with four convolutional blocks followed by a GRU layer, and 3) with seven convolutional blocks followed by a GRU layer.

\subsection{Comparison with Baselines}
\label{ssec:comp}

% \begin{figure*}[!htb]
%   \centering
%   \includegraphics[width=\linewidth]{WASPAA2020_PaperTemplate_LaTeX/figs/f1.pdf}
% \caption{Test macro f1 scores for the adaptive computation and baseline systems. The left figure shows the f1 against the maximum amount of input sub-bands the model observes; the right figure shows the f1 versus the number of MACs.}
% \label{fig:f1_multiclass}
% \end{figure*}

We first compare the frame-wise macro F1 score of the test data of the proposed adaptive frequency selection methods (HIDACT and ACT) to the baseline systems in \ref{ssec:expr_model}.
%Our preliminary experiments show that the original DACT, without interpolation in hidden state, fails to adapt to the test data with poor performance and we hence exclude the system from further discussion.
For each model, we make eight different test runs. In each run, we limit the maximum number of frequency sub-bands the model can observe during test time, from 1 to 8, to simulate an environment with a different budget on data and computation. We start with the lowest frequency sub-band and incorporate one contiguous sub-band at a time.

Figure~\ref{fig:f1_multiclass} (left) shows the curve of F1 versus the maximum number of input sub-bands the model takes as input. Notice that the f1 scores for the baseline models (SB, SBR, FB1, and FB3) always lie on an integer value of the x-axis, since these methods always process all available sub-bands. Among all the sub-band models, the HIDACT model on average processes the fewest frequency sub-bands (less than 6) while having the highest test F1. Both HIDACT and ACT achieve roughly 70\% F1 within an average of 5 computational steps per frame. The performance of the baseline sub-band model SB degrades significantly when not observing all 8 sub-bands. SBR, the baseline observing a random number of sub-bands during training, has a less significant performance drop for fewer steps, but the performance with all 8 sub-bands is significantly worse than the other systems. It is possible that SBR overfits an intermediate number of sub-bands and hence the performance stops improving as more input is available.

Figure~\ref{fig:f1_multiclass} (right) displays the test performance with MACs. Despite the overhead in estimating the uncertainty and the computational cost, these adaptive methods can reach better performance in fewer operations compared to the baseline sub-band SB and the baseline full-band FB1. The performance is comparable to the full-band model FB3 with $3\times3$ kernels but with much fewer parameters and only one-third of MACs.

\subsection{Visualization of Adaptive Frequency Selection}
\label{ssec:res_vis}

% TODO: adding event labels
% \begin{figure}[!htb]
%   \centering
%   \includegraphics[width=\linewidth]{WASPAA2020_PaperTemplate_LaTeX/figs/interpretation.pdf}
% \caption{Number of sub-bands processed by the HIDACT model for a test recording.}
% \label{fig:hidact_step}
% \end{figure}

We further analyze the behavior of our proposed adaptive computational method HIDACT with respect to the input. Figure~\ref{fig:hidact_step} shows the frame-wise number of frequency sub-bands processed by the HIDACT model, overlayed on the log mel-spectrogram with 128 features with 8 sub-bands of a test recording. Notice the lags of a few frames between the onset of the events and the change in the processing steps. When observing a new event, the network needs to process a few frames to leverage sufficient temporal information. On average, the model requires 6 sub-bands to assign a background label to the frames with no active events; for event classes \textit{coughing, knock}, and \textit{laughter} where the low-frequency components contain strong cues of the event types, the model processes 4-5 frequency sub-bands to make a confident prediction; in contrast, for the \textit{phone} class, most of the rich content and useful information is found at higher frequency bins, and hence the model takes up to 7 processing steps. This diagram indicates that HIDACT not only requires less input and computation but also adaptively adjusts the computation cost based on the input signal.

%% file: conclusion.tex
\section{Conclusion}
\label{sec:conclusion}

We have proposed a method to efficiently detect non-overlapping sound event classes with adaptive frequency selection and sub-band processing. The model is trained once and can adjust its computational cost according to the available computational budget during inference. It adapts to the input signal and learns to make confident predictions without processing all frequency sub-bands. Empirical results show that the proposed framework achieves comparable performance to full-band or sub-band models with higher computational complexity. In the future, we aim to extend the system to more complicated tasks including polyphonic event detection as well as more efficient architecture design.